\documentclass[floatfix,aps,twocolumn,preprintnumbers,amsmath,amssymb,nofootinbib,superscriptaddress,prl]{revtex4}

\usepackage{graphicx,color}
\usepackage{amsmath,amssymb,bm}
\usepackage{dcolumn}

\newcommand{\np}{{\bf p}}
\newcommand{\nq}{{\bf q}}

\def\XXint#1#2#3{{\setbox0=\hbox{$#1{#2#3}{\int}$}
     \vcenter{\hbox{$#2#3$}}\kern-.5\wd0}}

\def\1{\'{\i}}

\begin{document}

\title{Scaling violation and 
relativistic effective mass from quasielastic electron 
scattering: implications for neutrino reactions} 

\thanks{This
  work is supported by Spanish DGI (grant FIS2014-59386-P) and Junta de
  Andaluc{\'{\i}a} (grant FQM225). }

\author{J.E. Amaro}\email{amaro@ugr.es} \affiliation{Departamento de
  F\'{\i}sica At\'omica, Molecular y Nuclear \\ and Instituto Carlos I
  de F{\'\i}sica Te\'orica y Computacional \\ Universidad de Granada,
  E-18071 Granada, Spain.}
  
\author{E. Ruiz
  Arriola}\email{earriola@ugr.es} \affiliation{Departamento de
  F\'{\i}sica At\'omica, Molecular y Nuclear \\ and Instituto Carlos I
  de F{\'\i}sica Te\'orica y Computacional \\ Universidad de Granada,
  E-18071 Granada, Spain.} 

\author{I. Ruiz Simo}\email{ruizsig@ugr.es} \affiliation{Departamento de
  F\'{\i}sica At\'omica, Molecular y Nuclear \\ and Instituto Carlos I
  de F{\'\i}sica Te\'orica y Computacional \\ Universidad de Granada,
  E-18071 Granada, Spain.}

\date{\today}

\begin{abstract}
\rule{0ex}{3ex} The experimental data from quasielastic electron
scattering from $^{12}$C are reanalyzed in terms of a new scaling
variable suggested by the interacting relativistic Fermi gas with
scalar and vector interactions, which is known to generate a
relativistic effective mass for the interacting nucleons.  By choosing
a mean value of this relativistic effective mass $m_N^* =0.8 m_N$, we
observe that most of the data fall inside a region around the inverse
parabola-shaped universal scaling function of the relativistic Fermi
gas.  This suggests a method to select the subset of data that
highlight the quasielastic region, about two thirds of the total 2,500
data.  Regardless of the momentum and energy transfer, this method
automatically excludes the data that are not dominated by the
quasielastic process.  The resulting band of data reflects deviations
from the perfect universality, and can be used to characterize
experimentally the quasielastic peak, despite the manifest scaling
violation.  Moreover we show that the spread of the data around the
scaling function can be interpreted as genuine fluctuations of the
effective mass $M^* \equiv m^*_N/m_N \sim 0.8 \pm 0.1$.  Applying the same
procedure we transport the scaling quasielastic band into a
theoretical prediction band for neutrino scattering cross section that
is compatible with the recent measurements and slightly more
accurate.

\end{abstract}



\maketitle

Quasielastic electron scattering from nuclei has experienced a revival
out of the practical need to gauge the validity of the current models
as applied to neutrino scattering and oscillation experiments
\cite{Nomad09,Agu10,Agu13,Fio13,Abe13} (for recent reviews see
refs. \cite{Gal11,For12,Mor12,Alv14}).  The conventional approach
pursues a detailed microscopic relativistic description of the
inelastic processes and then requires all the relevant mechanisms for
the particular $Q^2$ kinematics. At present there is no compelling
model able to describe the world $(e,e')$ experiments. In the case of
$^{12}$C, taken as example here, the more than 2,500 data available
spread over a huge $(q,\omega)$ kinematical region, reaching well
inside the relativistic regime. A crucial issue is to find which
electron data encode the maximum information to be applied to neutrino
scattering minimizing the systematic and theoretical uncertainties in
the relevant channel (quasielastic, pion emission, \ldots).  The
scaling approach provides an appealing and unified framework to
encompass coherently the large diversity of data stemming from
different experiments and kinematics. In particular the super-scaling
approach (SuSA) has been implemented along these lines to predict
neutrino scattering cross sections from a longitudinal scaling
function $f_{L}(\psi')$ fitted to electron data
\cite{Ama04}. Moreover, the recent upgrade of the SuSA-v2 \cite{Gon14}
includes nuclear effects which are theoretically-inspired in a
particular realization of the relativistic mean field (RMF) theory, by
an additional transverse scaling function $f_T(\psi')$ which
 is different from $f_L(\psi')$.

A peculiar feature of the RMF is that it is the only approach which
reproduces the experimental scaling function $f_L(\psi')$ 
for all the values of $q$
after an {\em ad-hoc} $q$-dependent shift in energy is applied
\cite{Ama05}. The theoretical origin of this phenomenological shift
has not been well understood \cite{Gon14}.  This model incorporates a
dynamical enhancement of lower Dirac components, which is transmitted
to the transverse response, $R_T$, improving the agreement with experimental
data.  However, gauge invariance is violated  and hence
$R_T$ still presents ambiguities
\cite{Cab07}. Another difficulty of the RMF and other finite nuclei
models is that they break translational invariance (attempts to
restore it in a relativistic system were explored in
refs. \cite{Alb05,Alb07}).

The goal of this paper is to exploit the scaling idea from a novel
point of view connecting the RMF with the
universal scaling function of the relativistic Fermi gas (RFG)
\begin{equation}  f(\psi^*) = \frac{3}{4}(1-\psi^*{}^2) \theta(1-\psi^*{}^2) 
\label{f}
\end{equation}
Rather than constructing a yet undetermined scaling function we aim to
propose a new scaling variable $\psi^*$ mapping the data into a region
around the above function. Inspired by the fact that the mean field
theory provides a consistent and reasonable description of the nuclear
response in the quasielastic region (already observed by Rosenfelder
35 years ago \cite{Ros80}) for a range of kinematics, we propose to
start from the interacting RFG \cite{Ser86} including a suitable
vector and scalar potentials which are inferred from the data into an
effective mass $m^*_N$ that gets reduced in the nuclear medium.  The
effective mass encodes relativistic dynamical effects relevant in this
kinematical region, alternative to other approaches like the one based on
the spectral function \cite{Ben08,Ank15}.  In fact, one of the motivations
of our approach, called here $M^*$-scaling (or M*S), was to provide a
framework enjoying the good features of the RMF without incurring into
the above mentioned difficulties, unvealing the $m^*_N$ origin of 
the dynamical enhancement of both the lower components and  
 the transverse response function. The shift of $f_L(\psi')$
is trivially obtained 
as a consequence of the $m^*$ dependence of the quasielastic peak
position.

Of course, there have been numerous attempts to determine the
effective mass \cite{Weh93,Mariano05}, but this depends on details of
the dynamics. Thus, by proceeding directly from the data we avoid 
specifying the mean field explicitly. On the other hand a
phenomenological determination of $m_N^*$  suffers from the
uncertainties on the bulk of the data which should most
significantly contribute. Therefore, we will from the beginning accept
that this effective mass is determined up to a sensible uncertainty,
defined precisely by a suitable selection of the large database, to be
explained below in detail. One of the main advantages of this rather
simple approach, is not only its ease of implementation, but also that
we are free from the traditional objections regarding gauge invariance
or PCAC violations. We expect in this way to
account for the most relevant uncertainties regarding the predictive
power of the model.

We follow closely the notation introduced in Ref. \cite{Alb88}.
The quasielastic electroweak cross section is proportional to the
hadronic tensor or response function for single-nucleon excitations
transferring momentum $\nq$ and energy $\omega$, which in the Fermi
gas reads
\begin{eqnarray}
W^{\mu\nu}(q,\omega) 
&=& \frac{V}{(2\pi)^3} \int d^3p \delta(E'-E-\omega)
\frac{(m_N^*)^2}{EE'} 
\nonumber\\
&\times& 
2w^{\mu\nu}_{s.n.}(\np',\np)
\theta(k_F-p)\theta(p'-k_F)
\end{eqnarray}
where 
$E=\sqrt{\np^2+m_N^*{}^2}$ is the initial nucleon energy in the mean
field. The final momentum of the nucleon is $\np'=\np+\nq$ and its
energy is $E'=\sqrt{\np'{}^2+m_N^*{}^2}$. Note that initial and final
nucleons have the same effective mass $m_N^*$.  
The volume $V=3\pi^2 N/k_F^3$
of the system related to the Fermi momentum $k_F$ and proportional to
the number $N$ of protons and/or neutrons participating in the
process. Finally the electroweak interaction mechanism is implicit in the single-nucleon tensor
\begin{equation}
w^{\mu\nu}_{s.n.}(\np',\np)=\frac12\sum_{ss'}J^{\mu *}(\np',\np)J^{\nu}(\np',\np)
\end{equation}
where $J^{\mu *}$ is the electroweak current matrix element between
free positive energy Dirac spinors, with mass $m_N^*$ and
normalized to $\overline{u}u = 1$. In the case of electron scattering
we are involved with the electromagnetic current matrix element
\begin{equation}
J^\mu_{s's}(\np',\np)=
\overline{u}_{s'}(\np')
\left[ 
F_1(Q^2)\gamma^\mu 
+F_2(Q^2)i\sigma^{\mu\nu}\frac{Q_\nu}{2m_N}
\right]u_{s}(\np)
\end{equation}
where $F_1$ and $F_2$ are, respectively, 
 the Dirac and Pauli electromagnetic form factors of
proton or neutron.

In the case of $(e,e')$ the quasielastic cross section is written 
in Rosenbluth form
\begin{equation}
\frac{d\sigma}{d\Omega'd\epsilon'}
= \sigma_{\rm Mott}\left( v_L R_L + v_T R_T \right)
\end{equation}
where $\sigma_{\rm Mott}$ is the Mott cross section, $v_L= Q^4/q^4$
and $v_T=\tan^2(\theta/2)-Q^2/2q^2$, with $\theta$ the scattering
angle.  The nuclear longitudinal and transverse response functions are
the following components of the hadronic tensor in a coordinate sustem
with the $z$-axis in the $\nq$ direction (longitudinal)
\begin{eqnarray}
R_L(q,\omega) &=& W^{00} \\
R_T(q,\omega) &=& W^{11}+ W^{22}
\end{eqnarray}
In the RFG the nuclear response functions can be written in 
the  factorized form for $K=L,T$
\begin{eqnarray}
R_K & = &  G_K f(\psi^*),  \\
G_K &=& \Lambda(Z U^p_K+NU^n_K) 
\end{eqnarray}
Where $f(\psi^*)$ is given in Eq. (1) and $\psi^*$ is defined below. Moreover
\begin{equation}
\Lambda =  \frac{\xi_F}{m^*_N \eta_F^3 \kappa} 
\end{equation}
and the single nucleon response functions are
\begin{eqnarray}
U_L &=& \frac{\kappa^2}{\tau}
\left[ (G^*_E)^2 + \frac{(G_E^*)^2 + \tau (G_M^*)^2}{1+\tau}\Delta \right]
\\
U_T &=& 2\tau  (G_M^*)^2 + \frac{(G_E^*)^2 + \tau (G_M^*) ^2}{1+\tau}\Delta
\end{eqnarray}
where the quantity $\Delta$ has been introduced
\begin{equation}
\Delta= \frac{\tau}{\kappa^2}\xi_F(1-\psi^*{}^2)
\left[ \kappa\sqrt{1+\frac{1}{\tau}}+\frac{\xi_F}{3}(1-\psi^*{}^2)\right].
\end{equation}
Dimensionless variables have been introduced measuring the energy and
momentum in units of $m_N^*$, namely $\lambda= \omega/2m_N^*$,
$\kappa=q/2m_N^*$, $\tau=\kappa^2-\lambda^2$, $\eta_F= k_F/m_N^*$, and
$\xi_F=\sqrt{1+\eta_F^2}-1$. Note that usually \cite{Alb88} these
variables are defined with respect to the nucleon mass $m_N$ instead
of the $m_N^*$. The same can be said with respect to the electric 
and magnetic form
factors, that are modified in the medium due to the effective mass
according to 
\begin{eqnarray}
G_E^*  &=&  F_1-\tau \frac{m^*_N}{m_N} F_2 \\
G_M^*  &=& F_1+\frac{m_N^*}{m_N} F_2.
\end{eqnarray}
One should still stress that $F_1$ and $F_2$ can depend on $M^*$ \cite{Sai04}. 
We stick here to the phenomenologically succsessfull CC2 prescription
that reproduces the experimental superscaling function \cite{Cab07}.
Using the CC1 operator obtained through the Gordon reduction produces
the same effects as in the RMF of ref. \cite{Cab07}.  The same
modification of form factors in the medium was explored in
ref. \cite{Bar98}.
 For the free form factors we use the Galster
parametrization.

To define the scaling variable, $\psi^*$, we first introduce the minimum energy
allowed for a nucleon inside the nucleus to absorb the virtual photon
(in units of $m_N^*$)
\begin{equation}
\epsilon_0={\rm Max}
\left\{ 
       \kappa\sqrt{1+\frac{1}{\tau}}-\lambda, \epsilon_F-2\lambda
\right\}
\end{equation}
where $\epsilon_F=\sqrt{1+\eta_F^2}$ is the Fermi energy in units of $m_N^*$.
The scaling variable is defined by
\begin{equation}
\psi^* = \sqrt{\frac{\epsilon_0-1}{\epsilon_F-1}} {\rm sgn} (\lambda-\tau)
\end{equation}
Note that $\psi^* < 0 $ for $\lambda < \tau$ (the left side of the
quasielastic peak).
The meaning of $\psi^*{}^2$ is the following: it is the minimum kinetic 
energy of the initial nucleon divided by the kinetic Fermi energy.

Starting with the experimental $(e,e')$ cross section we compute
the experimental scaling function $f_{\rm exp}$ 
\begin{equation}
f_{\rm exp} = \frac{\left(\frac{d\sigma}{d\Omega'd\epsilon'}\right)_{\rm exp}}{
\sigma_{\rm Mott}\left( v_L G_L + v_T G_T \right)}
\end{equation}
which would correspond to the function $f(\psi^*)$ 
in the  relativistic Fermi
gas model.

\begin{figure}
\includegraphics[width= 8cm, bb=190 440 420 780]{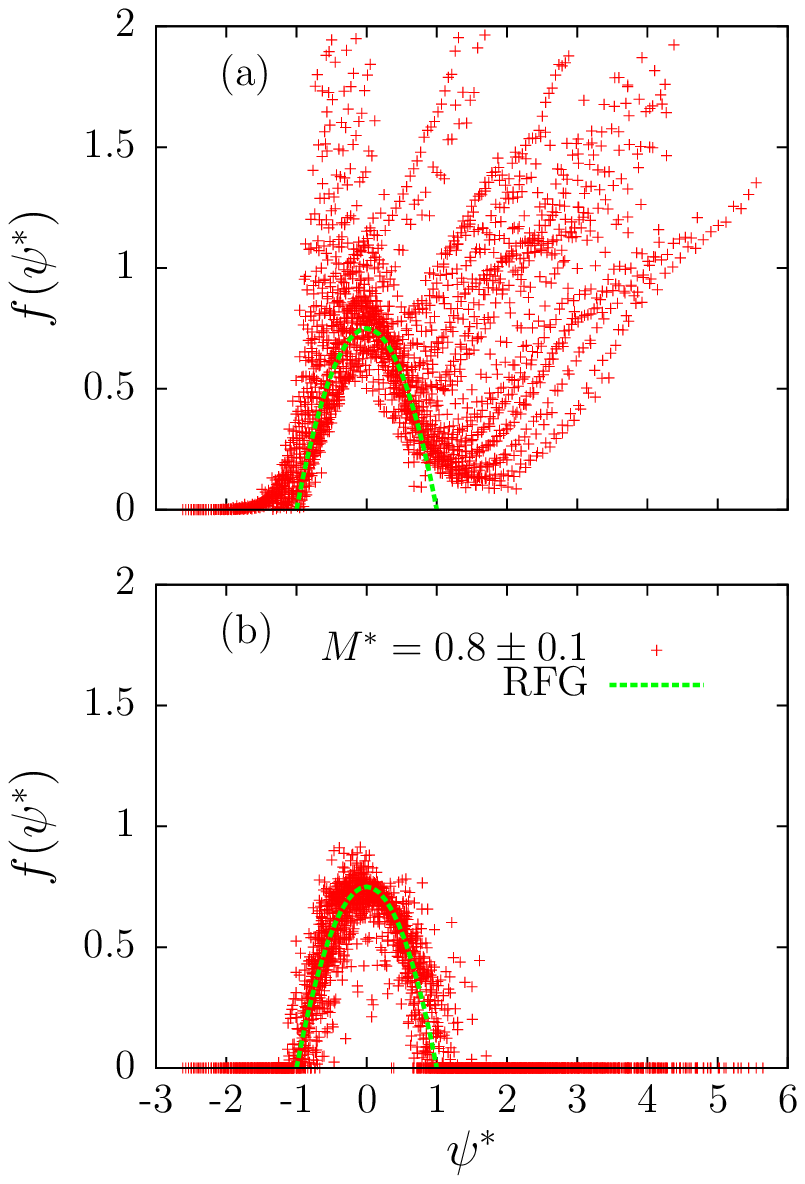}
\caption{Top panel (a): $M*$ scaling analysis of the experimental data
  of $^{12}$C as a function of the scaling variable $\psi^*$ for $M^*
  = m^*_N/m_N = 0.8$ compared to the RFG parabola. Bottom panel (b):
  RFG Monte Carlo simulation of QE data with a Gaussian distribution
  of relativistic effective mass quotient around $M^* = 0.8\pm
  0.1$. The Fermi momentum is fixed to $k_F=225$ MeV/c.  }
\end{figure}

We summarize the results of our M*S analysis in figure 1a.
We plot the experimental scaling function for the bulk of
$^{12}$C data \cite{archive,archive2} as a function of the scaling variable
$\psi^*$. We take 
\begin{equation}
M^* = \frac{m_N^*}{m_N} = 0.8.
\end{equation}
We see
that a large fraction of the data collapse into a data cloud
surrounding the RFG  scaling function, given by Eq
(\ref{f}).  Other choices of $m^*_N$ are possible but the clustering
substantially detunes from the RFG. So we interpret
this pattern as the kinematic regions highlighting the effective Fermi
gas behavior of data.  This collapse of data resolves two issues
simultaneously: On the theoretical side it provides an operational
definition of the relativistic effective mass, whereas on the
experimental side provides an operational definition of the
quasielastic peak behavior.

\begin{figure}
\includegraphics[width= 8cm, bb=130 440 440 780]{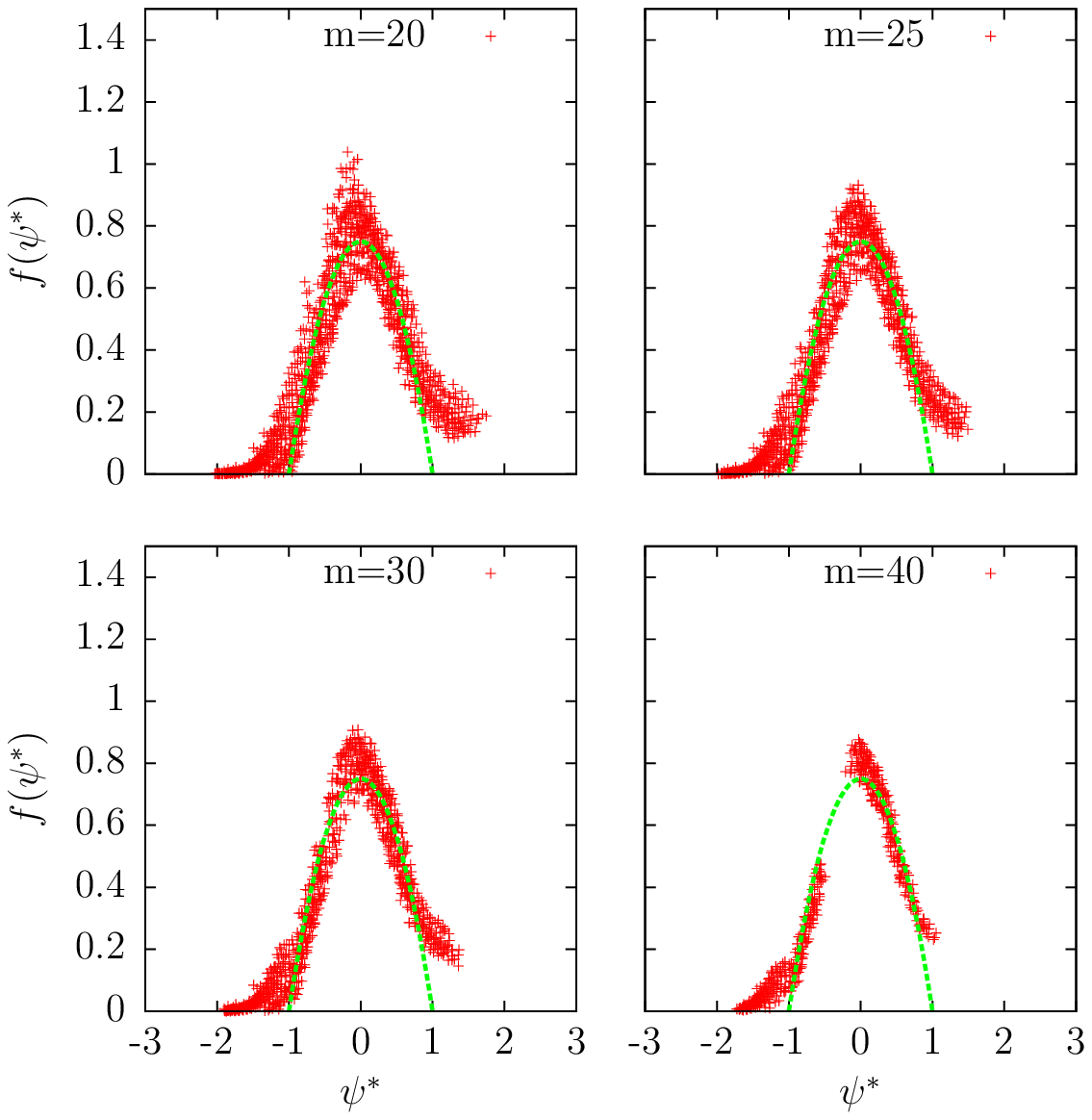}
\caption{
Experimental data selection in terms of the
  scaling variable $\psi^*$, obtained with different choices of the 
number $m$ of points inside a circle with radius $r=0.1$. 
}
\end{figure}

The observed scaling is not perfect in the sense that the blur of data
presents a finite width, but the width is roughly homogeneous as seen
in Fig. 2. There we select the data that are clustered on a coarse
grain scale according to a method inspired by the visual and
conventional Gaussian low-pass filtering (Gaussian blur) \cite{Gauss}.
Due to the discrete, heterogeneous and finite nature of the data in
our case we use instead a constant weighting function. This function
measures the density of points clustered above a given threshold $m$,
inside a circle of radius $r$ centered at the experimental point, plus
minus the experimental error. In the figure we show four situations
corresponding to $r=0.1$, and for illustration the result of applying
our low pass filtering method to four values of $m=20, 25, 30$ and 40.
The parameter $m$ measures the minimum number of experimental points
surrounding each data in the cloud. Note that we discard the
surrounding points that don't verify the above condition. As we can
see the shape defined by the data cloud, seen as a shaded band in the
scale of the figure, presents a stable pattern around the relativistic
Fermi gas when the threshold value increases, even if the number of
surviving points decreases. This stability around the Fermi gas result
is triggered by the chosen value of $M^*$. Note that the
number of data involved in these plots is around 1,500, but the
scaling violation (defined as the width of the shaded band) is
manifest. 

This pattern in the M*S plot, that emerges as a
realization of an universal quasielastic peak, is a global property of
the set of data and suggests an alternative interpretation in terms of
fluctuations of $M^*$. One could propose a statistical
model where each point in the cloud samples a quasielastic event with
a slightly different effective mass around the mean value 0.8.  This
fluctuation does not simulate the nuclear effects beyond the impulse 
approximation
(finite size effects, short-range NN-correlations, long-range RPA,
meson-exchange currents, $\Delta$ excitation, pion emission,
two-particle emission, final state interaction). However the fluctuations are 
 of the same order of magnitude as  these effects.
Actually they are small enough to
retain the points in the neighborhood of the quasielastic region,
which could be treated perturbatively in a microscopic framework
beyond the RFG. The largest deviations of the quasielastic cloud from
the perfect parabola occur only around its edges, where the Fermi gas
is zero and hence the resulting signal cannot be accounted for by a
change of $m_N^*$.

\begin{figure}
\includegraphics[width= 8cm, bb=170 440 450 780]{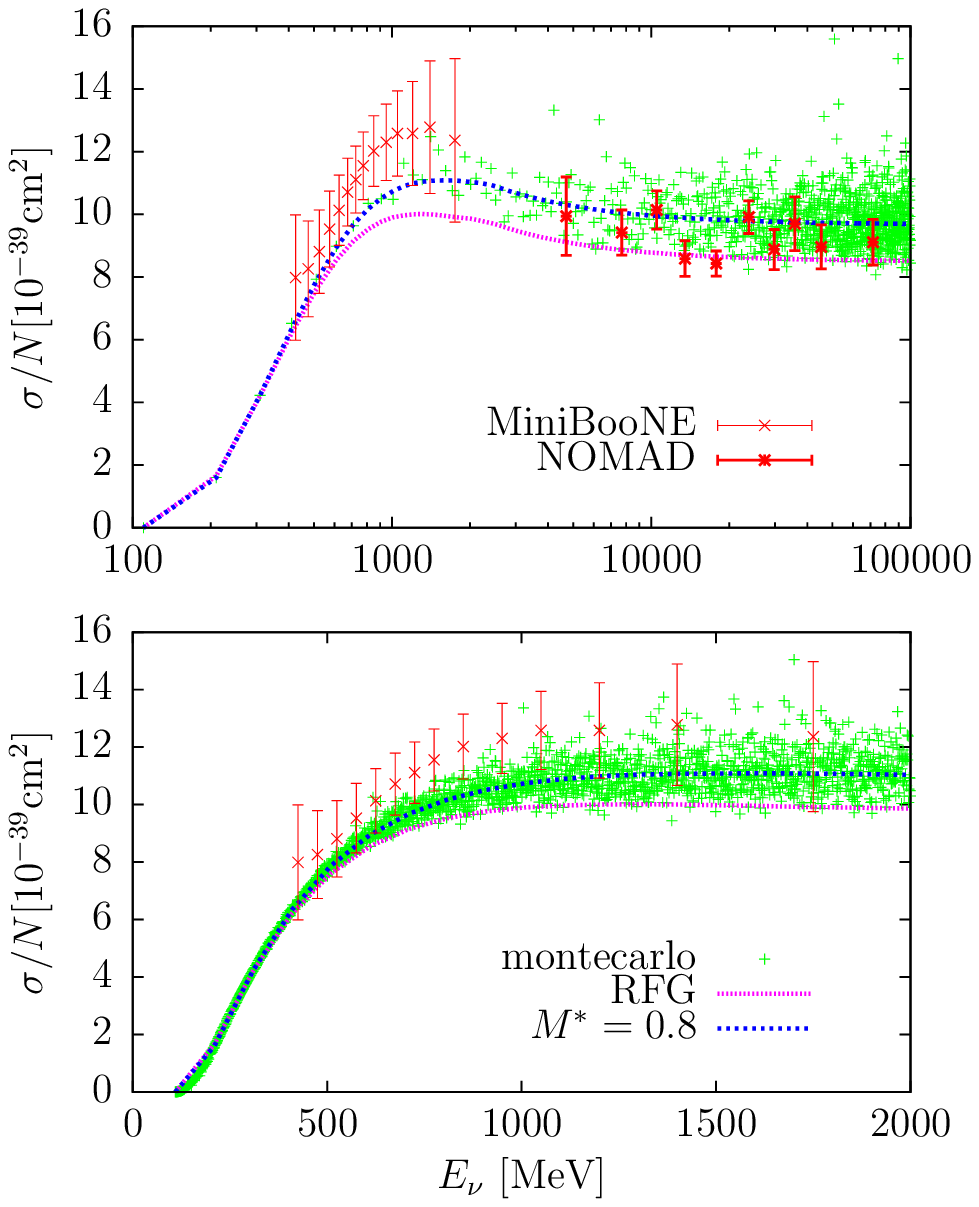}
\caption{ Total QE neutrino cross section off $^{12}$C per neutron as
  a function of the neutrino energy for different relativistic
  effective masses generated in a Monte Carlo simulation.  The
  experimental data points are from NOMAD \cite{Nomad09} and MiniBooNE
  \cite{Agu10}. We take the axial dipole mass $M_A=1$ GeV. }
\end{figure}

To justify the above assumption, we carried out a calculation using a
family of RFGs with slightly different $m^*_N$,
to generate a random point for each single experimental datum at the
very same kinematics.  Thus we take a random $M^*$  around the
optimal mean value 0.8 in a Monte Carlo sampling. The results of this
simulation are shown in Fig.1-b. To generate the pseudo data we use a
Gaussian distribution with a width value $\sigma= 0.1$, representing
the fluctuation of $M^* = 0.8 \pm 0.1$, that nicely
resembles the fluctuations seen in the cloud of the experimental
data. This procedure automatically selects those pseudo data
attributable to genuine quasielastic interpretation (based on the
Fermi gas definition) and makes zero those kinematics that are
forbidden. 

Our main observation is that by choosing the optimum relativistic
effective mass, a  RFG-like scaling of the data 
can be obtained in the quasielastic region, covering more than 1500
data. This implies a trade off between the experimental uncertainty to
what extent a datum is close to quasielastic and the importance of the
physical effects beyond the impulse approximation that contribute to
the quasielastic mechanism.  With this procedure a way to estimate what
information is contained in the data about the quasielastic peak
emerges.

From our analysis a phenomenological scaling function could be also
obtained exactly in the same way as in the superscaling analysis. We
have not tried to parameterize this function, that could be done from
the data of fig. 2. The resulting scaling function is asymmetrical and
very similar to the longitudinal superscaling function, but with a
different normalization, including the tail. Therefore the tail of the 
scaling function is a property of the quasielastic interaction.

The information extracted here about the quasielastic $(e,e')$ cross
section of $^{12}$C can be straightforwardly used to make readily
predictions for other reactions like CC neutrino scattering from the
same nucleus.  In figure 3 we show the calculations of the
$(\nu_{\mu},\mu^-)$ cross section as a function of incident neutrino
energy.  There we show the effective RFG results with $m_N^*= 0.8
m_N$. The cloud of points correspond to incorporating the same
fluctuations $\pm 0.1$ of $M^*$ as in the Monte Carlo simulation
depicted in Fig. 1-b. By comparison we also show the results of the
conventional RFG. The effective mass produces an enhancement of the
lower Dirac components, and hence also of both the vector and axial
transverse responses, and of the theoretical cross section, which
thanks to the fluctuations becomes compatible with the data for all
the kinematics. In our case the fluctuations of the theoretical band
are about 10\%, as naively expected from the input uncertainty of the
effective mass.  We note that the sampling in the lower panel of
figure 3 uses a smaller binning of 1 MeV as opposed to the upper panel
where 100 MeV was used instead. The clustering of these equidistant
binnings arises naturally from the log scale.

For high $Q$ the vector form factors deviate from the conventional
dipole behaviour \cite{Bod08}, which could affect, in principle, any
model's predictions. However this would only be appreciable in the
differential cross section; the integrated $\sigma$, even for NOMAD
kinematics, is only sensitive to the kinematical regions where the
product of form factors and phase space is large. We have numerically
checked that the contribution from $Q^2$ above 1-2 (GeV/c)$^2$ is
negligible, because of the rapid fall of the nucleon form factors.
As a matter of fact one can ignore the electric neutron form factor
completely after integration.

Note that the set of data of unfolded energy dependent CCQE cross
section model suffer from uncertainties driven by the model dependence
of the neutrino energy reconstruction.  The comparison of fig. 3 is
merely indicative for illustration purposes of the kind of predictions
that the present approach can provide for proper flux-averaged doubly
differential cross sections.  These comparison will be presented in a
forthcoming publication.

At present there is no model able to reproduce the 2,500 data points
from $^{12}C(e,e')$ experiments. Due to the impossibility to fit the
quasielastic peak or other regions with the experimental accuracy, in
the present approach, we have shown that instead of making an
extremely detailed analysis of the particular reaction, which maybe
well beyond the present validation possibilities, it is possible to
isolate those data contributing to the simplest possible physics
we are interested in, and use that information to make
predictions with the maximum allowed precision, since one cannot
distinguish the theoretical noise from the experimental signal.


\end{document}